\documentclass[aps,prl,twocolumn,superscriptaddress,showpacs]{revtex4}
\usepackage{graphicx}
\usepackage{epsfig}
\usepackage{amsmath}
\usepackage{amssymb}
\usepackage{amsfonts}
\usepackage{mathrsfs}
\usepackage{theorem}
\usepackage{bm}
\usepackage{url}
\usepackage[T1]{fontenc}
\usepackage{csquotes}
\usepackage{natbib}
\usepackage{float}
\usepackage{appendix}

\MakeOuterQuote{"}

\usepackage{dcolumn}
\usepackage{color}
\usepackage{color}

\begin{document}

\title{Heisenberg's error-disturbance relations: a joint measurement-based experimental test}

\author{Yuan-yuan Zhao}
\affiliation{Key Laboratory of Quantum Information, University of Science and Technology of China, CAS, Hefei, 230026, People's Republic of China}
\affiliation{Synergetic Innovation Center of Quantum Information and Quantum Physics, University of Science and Technology of China, Hefei, Anhui 230026, People's Republic of China}
\author{Pawe{\l} Kurzy\'{n}ski}
\email{cqtpkk@nus.edu.sg}
\affiliation{Centre for Quantum Technologies, National University of Singapore, 3 Science Drive 2, 117543 Singapore, Singapore}
\affiliation{Faculty of Physics, Adam Mickiewicz University, Umultowska 85, 61-614 Pozna\'{n} , Poland}
\author{Guo-yong Xiang}
\email{gyxiang@ustc.edu.cn}
\affiliation{Key Laboratory of Quantum Information, University of Science and Technology of China, CAS, Hefei, 230026, People's Republic of China}
\affiliation{Synergetic Innovation Center of Quantum Information and Quantum Physics, University of Science and Technology of China, Hefei, Anhui 230026, People's Republic of China}
\author{Chuan-Feng Li}
\affiliation{Key Laboratory of Quantum Information, University of Science and Technology of China, CAS, Hefei, 230026, People's Republic of China}
\affiliation{Synergetic Innovation Center of Quantum Information and Quantum Physics, University of Science and Technology of China, Hefei, Anhui 230026, People's Republic of China}

\author{Guang-Can Guo}
\affiliation{Key Laboratory of Quantum Information, University of Science and Technology of China, CAS, Hefei, 230026, People's Republic of China}
\affiliation{Synergetic Innovation Center of Quantum Information and Quantum Physics, University of Science and Technology of China, Hefei, Anhui 230026, People's Republic of China}

\begin{abstract}
The Heisenberg's error-disturbance relation is a cornerstone of quantum physics. It was recently shown to be not universally valid and two different approaches to reformulate it were proposed \cite{Ozawa1,Ozawa2,Branciard,BLWpq,BLWQ}. The first one focuses on how error and disturbance of two observables, A and B, depend on a particular quantum state \cite{Ozawa1,Ozawa2,Branciard}. The second one asks how a joint measurement of A and B affects their eigenstates \cite{BLWpq,BLWQ}. Previous experiments focused on the first approach \cite{Erhart,Rozema,BKOE,Weston,RBBFBW,KBOE,Entropic}. Here, we focus on the second one. Firstly, we propose and implement an extendible method for quantum walk-based joint measurements of noisy Pauli operators to test the error-disturbance relation for qubits introduced in Ref. \cite{BLWQ}. Then, we formulate and experimentally test a new universally valid relation for the three mutually unbiased observables. We therefore establish a fundamentally new method of testing error-disturbance relations.
\end{abstract}
\maketitle

The test of Heisenberg's error-disturbance relation has been recently proposed \cite{Ozawa1,Ozawa2,LW} and implemented \cite{Erhart,Rozema,BKOE,Weston,RBBFBW,KBOE,Entropic} and the observed violation became a subject of a heated debate \cite{Ozawa1,Ozawa2,Branciard,BLWpq,BLWQ,BLW,BS}. The relation can be formulated as $\varepsilon(A)\eta(B)\geq \frac{1}{2} |\langle [A,B] \rangle$, were $\varepsilon(A)$ is the measurement precision of an observable $A$ and $\eta(B)$ is the disturbance that this measurement induces on another observable $B$. The violation stems from states for which measures of error and disturbance can be both zero $\varepsilon(A)=\eta(B)=0$, while at the same time the estimated lower bound on them is non-zero \cite{Branciard,BLW,Rudolph}. This motivates a state-independent approach \cite{BLWpq,BLWQ} in which the error-disturbance trade-off is evaluated for a set of states that "calibrate" the measurement apparatus \cite{BLWpq}.

\begin{figure}[t]
    \begin{center}
    	\includegraphics[width=1.0\columnwidth,trim=4 4 4 4 ,clip]{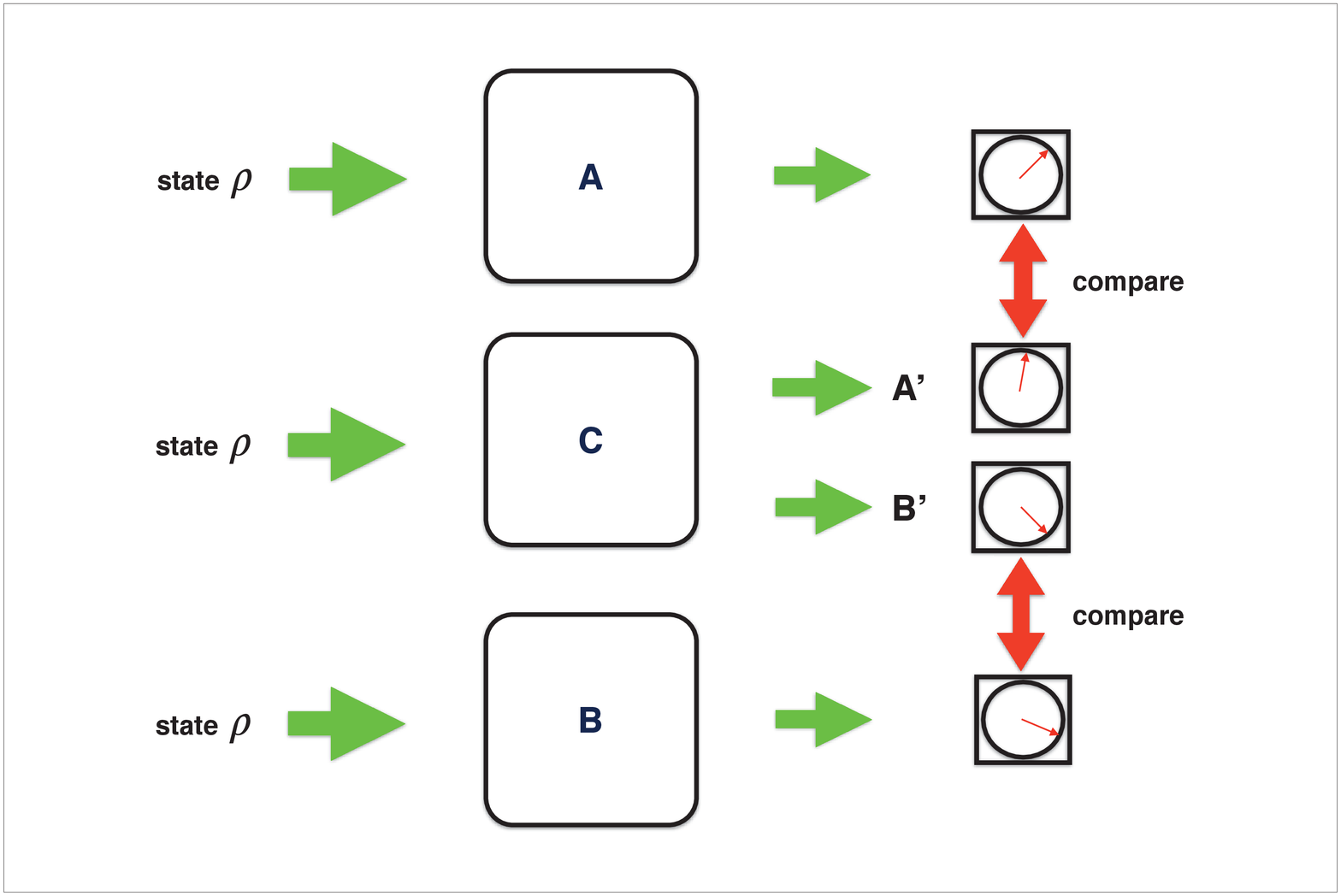}
       \vspace{-5mm}
        \caption{Joint measurement of A and B cannot be done with perfect precision due to their incompatibility. Instead, one looks for an optimal observable $C$ from which noisy versions of the two observables, $A'$ and $B'$, can be estimated. The error-disturbance trade-off is evaluated by a comparison of the outcomes of $A'$ ($B'$) with the outcomes of $A$ ($B$) when measured alone.}
        \label{fig1}
    \end{center}
\end{figure}

More precisely, the idea of the state-independent approach \cite{BLWpq,BLWQ} is to test the measurement-disturbance tradeoff in a setup allowing for a joint measurement of both observables, see Fig. \ref{fig1}. In general, an exact measurement of $A$ and $B$ cannot be performed if the two observables are incompatible. However, one can find another measurement procedure, let's call it $C$, whose outcomes can be used to estimate $A'$ and $B'$ that are noisy versions of the original observables. The error-disturbance trade-off can be later evaluated by comparing the outcomes of $A'$ and $B'$ with the outcomes of $A$ and $B$ that were obtained in scenarios when the original observables were measured separately.

\begin{figure*}
\begin{center}
    	\includegraphics[width=1.8\columnwidth,height=9.5cm]{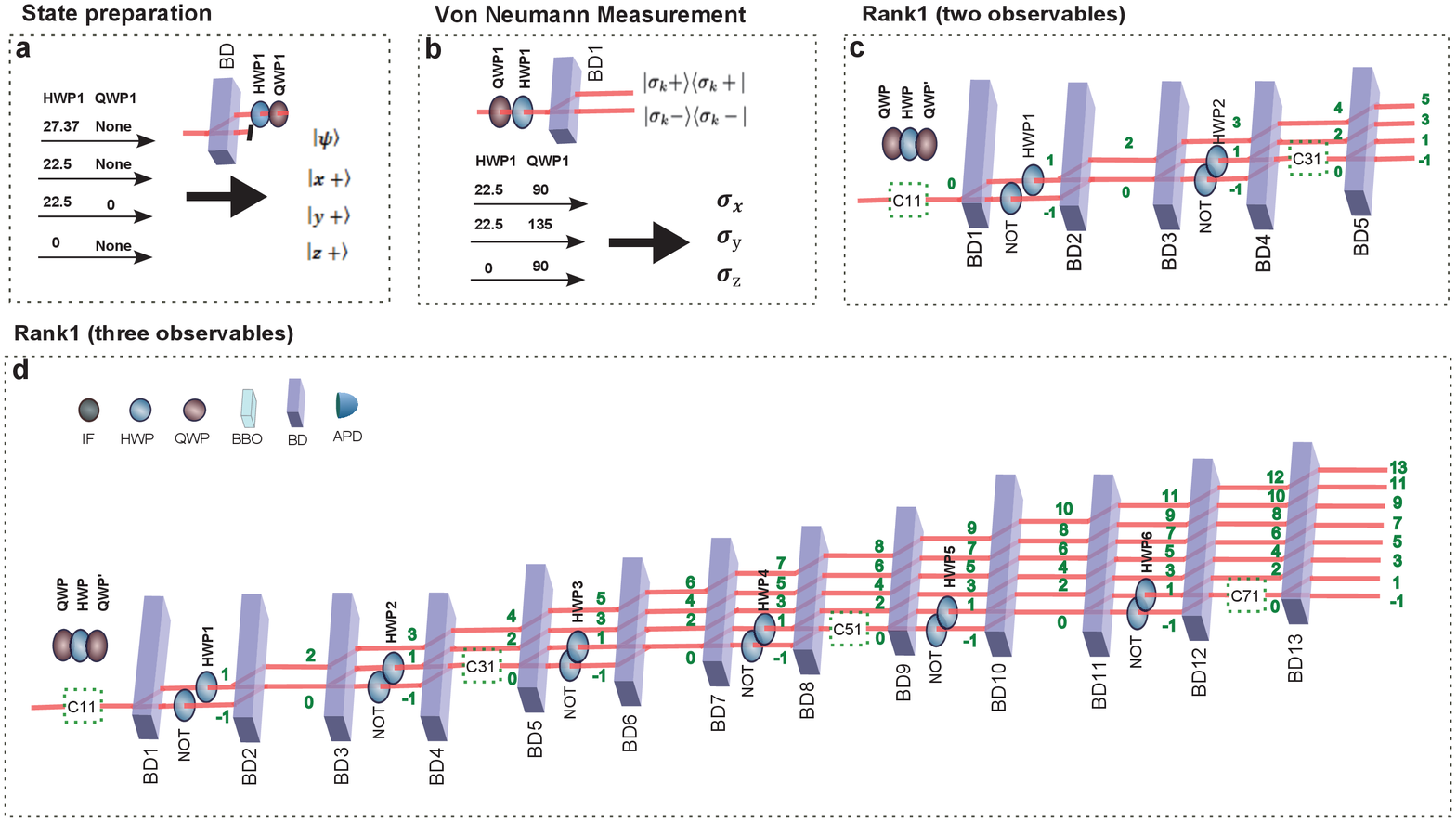}

        \caption{{\bf Experimental setup.} {\bf a,} The configuration for preparing four input states and the relevant settings of HWP1 and QWP1. {\bf b,} Setup for conducting the Von Neumann Measurement of $\sigma_x$, $\sigma_y$ and $\sigma_z$ and the relevant settings of HWP1 and QWP1. $|\sigma_k+\rangle\langle\sigma_k+|$ is the measurement operator corresponding to the outcome $\sigma_k=1$, where k=x, y and z. {\bf c,} Optical network for Joint measurement of three pairs of Pauli operators: ($\sigma_x$, $\sigma_y$), ($\sigma_x$, $\sigma_z$) and ($\sigma_y$, $\sigma_z$). {\bf d,} Optical network for joint measurement of three Pauli operators. }\label{fig2}
    \end{center}
\end{figure*}

Instead of using $\varepsilon(A)$ and $\eta(B)$ it was proposed to measure the error-disturbance trade-off in terms of a distance $\Delta(A,A')$ and $\Delta(B,B')$ \cite{BLWpq}. Intuitively, $\Delta(A,A')$ measures how well $A'$ approximates $A$ (similar for $B$ and $B'$). Next, it was shown in Ref. \cite{BLWQ} that for $\pm 1$ qubit measurements the state-dependent distance is of the form
\begin{equation}
\Delta(A_{\rho},A'_{\rho})^2 = 4\left|p(A+)-p(A'+)\right|,
\end{equation}
where $\rho=\frac{1}{2}(\openone + \mathbf{r} \cdot \mathbf{\sigma})$, $p(A+)=Tr\{\rho E(A+)\}$ and $E(A+)$ is the generalised measurement operator (Positive Operator Value Measure -- POVM) corresponding to the outcome $A=+1$ -- for details see Methods. The qubit POVM operators can be expressed as $E(A+) = \frac{1}{2}(a_0 \openone + \mathbf{a} \cdot \mathbf{\sigma})$ and $E(A'+) = \frac{1}{2}(a'_0 \openone + \mathbf{a'} \cdot \mathbf{\sigma})$, therefore $\Delta(A_{\rho},A'_{\rho})^2 = 2|a_0 - a'_0 + \mathbf{r\cdot(a-a')}|$. In order to obtain state-independent distance one chooses $\mathbf{r}$ to be a unit vector lying along $\mathbf{a-a'}$ and therefore $\Delta(A,A')^2=2|a_0-a'_0| + 2\| \mathbf{a-a'} \|$. In particular, if $A$ is a sharp observable then $a_0=1$ and $\mathbf{a}$ is a unit vector. In addition, if we choose $A'$ to be an unsharp version of $A$ such that
\begin{equation}
E(A'+) = \frac{1}{2}( \openone + \eta \mathbf{a} \cdot \mathbf{\sigma}),
\end{equation}
where $0 < \eta \leq 1$ is the efficiency, then the distance is maximised for $\mathbf{r}=\mathbf{a}$. Using the above choice we can think of the measurement process as of "calibration" \cite{BLWpq}, since for such states the outcome of $A$ is known. In this case $\Delta(A,A')^2=2(1-\eta)$.

The Heisenberg-type relation for qubit measurements that uses the distance approach is of the form \cite{BLWQ}
\begin{equation}\label{rel}
\Delta(A,A')^2 + \Delta(B,B')^2 \geq \sqrt{2} \left( \| \mathbf{a} - \mathbf{b} \| + \| \mathbf{a} + \mathbf{b} \| -2 \right).
\end{equation}
The maximal value of the lower bound is $4-2\sqrt{2}$ and can be obtained for mutually unbiassed observables in which case $\mathbf{a} \perp \mathbf{b}$. The equality is achieved for $\eta = 1/\sqrt{2}$. Interestingly, joint measurement of two unsharp unbiased observables is possible for $\eta \leq 1/\sqrt{2}$ \cite{HRS,SRH,Busch2,Yu}.

In Ref. \cite{QWPOVM} it was proposed that POVM measurements can be naturally implemented in a discrete-time quantum walk setup. This idea was recently experimentally verified \cite{Us,Xue}. Such setup perfectly matches the joint-measurement requirements discussed above, since it naturally implements a measurement $C$ that allows to estimate $A'$ and $B'$ which are approximations of $A$ and $B$. Moreover, substantial control over the quantum walk parameters allows one to implement joint measurements of practically any pair of qubit measurements and even of a triple of qubit measurements.

A discrete-time quantum walk is a simple model of a single particle dynamics moving in a discrete space. Here, we consider a simple case of a one-dimensional space \cite{Aharonov}. The state of the particle is described by $|x,c\rangle$, where $x=\ldots,-1,0,1,\ldots$ is the position and $c=\leftarrow,\rightarrow$ is a two-level degree of freedom known as a coin. In each step of the evolution the particle moves either to the left or right, depending on the state of the coin. This conditional translation is described by
\begin{equation}\label{T}
T=\sum_x |x+1,\rightarrow\rangle \langle x,\rightarrow | + |x-1,\leftarrow \rangle \langle x,\leftarrow |.
\end{equation}
In addition, the coin degree of freedom also evolves according to
\begin{equation}\label{C}
C=\begin{pmatrix} \cos\theta & e^{-i\beta}\sin\theta \\ -e^{i\beta}\sin\theta & \cos\theta \end{pmatrix}
\end{equation}
in order to allow for a superposition of walking right and left at the same time. A single step of the evolution is given by $TC$ and it can be considered as a standard von Neumann measurement of the coin. If the walker moved to the right we know that the coin was initially in the state $C^{\dagger}|\rightarrow\rangle$ and if it moved to the left we know that the coin was in the state $C^{\dagger}|\leftarrow\rangle$.

However, we can consider more than one step of the underlying quantum walk in which the operator $C$ can depend on time and position. This allows for a nontrivial engineering of correlations between the position and the coin. As a result, the measurement of the particle's position constitutes a POVM on a coin degree of freedom. Then we construct the joint measurements of two and three observables based on quantum walk, and use our joint measurements to test the error-disturbance relation for qubits.


%
%
{\bf Experimental setup.}
For the quantum walk setup we constructed, the position of the particle $x$ and the coin state $c$ are encoded in the longitudinal spatial modes and polarizations {$|H\rangle$, $|V\rangle$} of the single photons. The conditional translation operator $T$ is realized by the designed BD, that does not displace the vertical polarized photons ($|x, V\rangle\rightarrow|x, V\rangle$) but makes the horizontal polarized ones undergo a 4 mm lateral displacement ($|x, H\rangle\rightarrow|x+1, H\rangle$). In the experiment, the input states are prepared by passing single photons through a BD (the vertical polarized photons are blocked), a half-wave plate (HWP) and a quarter-wave plate (QWP) in a specific configuration. The detailed settings can be found in Fig. \ref{fig2}.

\begin{figure}[htbp]
\begin{center}
    	\includegraphics[width=1\columnwidth]{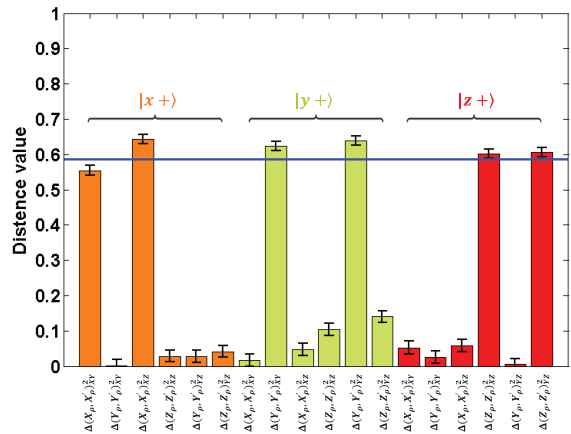}
        \caption{{\bf Distance values for joint measurement of two unsharp Pauli observalbles.} Different colour bars represent the different input states: $|x+\rangle$, $|y+\rangle$ and $|z+\rangle$. The values of state-independent calibration distances are shown as the blue lines. The error bars are obtained by Monte Carlo simulation (1,000 runs)}
    \label{result2}
    \end{center}
\end{figure}

{\bf Experimental test of the Busch-Lahti-Werner qubit relation.}
We consider joint measurements of noisy pairs $\{X',Y'\}$, $\{Y',Z'\}$ and $\{X',Z'\}$. We aim to achieve the efficiency $\eta = 1/\sqrt{2}$ (see Supplementary Information for detailed results). In addition, we perform standard separate von Neumann measurements of $X$, $Y$ and $Z$. This is necessary to evaluate the state-dependent and state-independent distances. The measurement of calibration states allows us to test (\ref{rel}) for all possible pairs, as well as to measure state-dependent  distances $\Delta(A_{\rho},A'_{\rho})^2$ and to observe that in some cases they can be very small. The measured distances are shown in Fig. \ref{result2}.

Next, we identify the state-independent calibration distances. Their theoretical value is $2(1-1/\sqrt{2}) \approx 0.5858$. We get $\Delta(X,X')_{XY}^2=0.5545\pm0.0134$ and $\Delta(Y,Y')_{XY}^2=0.6235\pm0.0132$, $\Delta(X,X')_{XZ}^2=0.6425\pm0.0128$ and $\Delta(Z,Z')_{XZ}^2=0.6011\pm0.0133$, $\Delta(Y,Y')_{YZ}^2=0.6386\pm0.0129$ and $\Delta(Z,Z')_{YZ}^2=0.6055\pm0.0133$.

The sums of the pairs of calibration distances are $1.1780\pm0.0266$, $1.2436\pm0.0261$ and $1.2441\pm0.0262$, respectively. Because we measure mutually unbiassed observables the lower bound of (\ref{rel}) is $4-2\sqrt{2} \approx 1.1716$. We therefore confirm the validity of (\ref{rel}). We also observed that for specific states some pairs of state-dependent distances are small. For example, $\Delta(Y_{x+},Y'_{x+})_{YZ}^2=0.0277\pm0.0173$ and $\Delta(Z_{x+},Z'_{x+})_{YZ}^2=0.0414\pm0.0169$. A similar effect lies at the root of the violation of state-dependent error-disturbance relations. Moreover, this result confirms observations in \cite{Branciard,BLW,Rudolph}. Our measurement data also allows us to reconfirm some of the previous experimental results \cite{Erhart,Rozema,BKOE,Weston,RBBFBW,KBOE} -- see supplemental material.


{\bf New relation for three unbiassed observables. }
We propose a special scenario in which one can use state-dependent error-disturbance measures by adding an additional measurement to the relation. In particular, we consider a trade-off relation between the three unbiased Pauli observables. In this case the measurements span all three directions in the Bloch-ball representation of the qubit and therefore the state of the system does not stand out from the parameter space described by the measurement.

Let us consider a joint measurement of the three unsharp Pauli observables $X'$, $Y'$ and $Z'$, with the corresponding efficiencies $\eta_x$, $\eta_y$ and $\eta_z$, that approximate the original spin operators. For the state $\rho$ one has
\begin{eqnarray}
& & \Delta(X_{\rho},X'_{\rho})^2 + \Delta(Y_{\rho},Y'_{\rho})^2 + \Delta(Z_{\rho},Z'_{\rho})^2    \label{rel2} \\
&\geq& r \min\{ \Delta(X,X')^2,\Delta(Y,Y')^2,\Delta(Z,Z')^2\}. \nonumber
\end{eqnarray}
For derivation see -- methods.

\begin{figure}[htbp]
\begin{center}
    	\includegraphics[width=0.98\columnwidth,height=10cm]{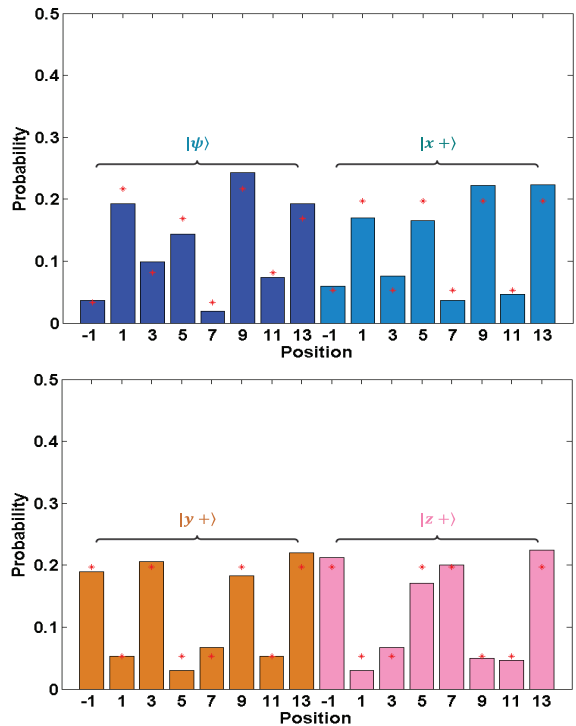}
        \caption{{\bf Normalized probability distributions for joint measurement of three noisy observables X, Y and Z.} Different colour bars represent the different input states: $|\psi\rangle$, $|x+\rangle$, $|y+\rangle$ and $|z+\rangle$. The theory values are shown as the star markers, and error bars are too small to identify. }
    \label{result3}
    \end{center}
\end{figure}

Next, we experimentally test the above relation. We perform a joint measurement of unsharp versions of $X$, $Y$ and $Z$ (results in Fig. ~\ref{result3}). In particular, we focus on the case $\eta_x=\eta_y=\eta_z=\eta$ for which the three unsharp observables are jointly measurable if $\eta \leq 1/\sqrt{3}$ \cite{HRS,SRH,Busch2,Yu}. We aim at $\eta=1/\sqrt{3}$.

We obtained the following state-independent distances: $\Delta(X,X')^2=0.8728\pm0.0211$, $\Delta(Y,Y')^2=0.8064\pm0.0217$, $\Delta(Z,Z')^2=0.7681\pm0.0216$. The theoretical value for all three distances is $2(\sqrt{3}-1)/\sqrt{3}\approx 0.845$.

Next, we evaluated state-dependent distances. For state $|x+\rangle$ we obtained $\Delta(X_{x+},X'_{x+})^2=0.8728\pm0.0211$, $\Delta(Y_{x+},Y'_{x+})^2=0.2984\pm0.0240$, and $\Delta(Z_{x+},Z'_{x+})^2=0.0828\pm0.0222$. For $|y+\rangle$ we obtained $\Delta(X_{y+},X'_{y+})^2=0.1176\pm0.0227$, $\Delta(Y_{y+},Y'_{y+})^2=0.8064\pm0.0217$, and $\Delta(Z_{y+},Z'_{y+})^2=0.0611\pm0.0229$. Finally, for $|z+\rangle$ we obtained $\Delta(X_{z+},X'_{z+})^2=0.1184\pm0.0226$, $\Delta(Y_{z+},Y'_{z+})^2=0.1934\pm0.0237$, and $\Delta(Z_{z+},Z'_{z+})^2=0.7681\pm0.0216$.

The above values can be plugged to relation (\ref{rel2}), however due to experimental noise they do not saturate it. The relation is more interesting when tested for state $|\psi\rangle$. In this case we obtain $\Delta(X_{\psi},X'_{\psi})^2=0.7932\pm0.0233$, $\Delta(Y_{\psi},Y'_{\psi})^2=0.2652\pm0.0238$, and $\Delta(Z_{\psi},Z'_{\psi})^2=0.1932\pm0.0201$. In addition, to evaluate the lower bound we need to know the value of $r$. Although in principle the state was assumed to be pure, the actual value of $r$ can be evaluated from the average values of von Neumann measurements $\sqrt{\langle X\rangle^2 + \langle Y\rangle^2 + \langle Z\rangle^2} = 0.9888$. Plugging all this to (\ref{rel2}) one gets $1.2516\pm0.0672 > 0.7595$.

To conclude, we report an experimental implementation of quantum walk-based joint measurements of noisy Pauli operators. We use the optical setup in which the spatial modes of a single photon correspond to the walker's position and the polarisation plays the role of the coin. Next, we use the outcomes from a joint measurement of two observables to test the error-disturbance relation for qubits introduced in Ref. \cite{BLWQ}. Note, that the previous experiments testing error-disturbance relations were based on the weak-value method or the so-called three-state method \cite{BS} and that the quantum walk based joint measurement method used by us is fundamentally different. Finally, we propose a modified relation for three mutually unbiassed observables and test it using the outcomes from a joint measurement of all three Pauli operators.


\section{Appendix}

{\bf Generalised quantum measurements.}
In this work we consider polarisation measurements of a single photon, that can be represented by binary $\pm 1$ observables. A joint measurement of two such observables requires four outcomes which can be obtained with a help of Positive-Operator Valued Measure (POVM). The POVM elements $E_i$ ($i=1,2,\dots,n$) obey
\begin{equation}
\forall_{\rho} ~\text{Tr}\{\rho E_i\} \geq 0,~~~~\sum_i^n E_i = \openone.
\end{equation}
They do not form an orthonormal set and hence their number can be greater than the dimension of the system. In order to implement POVMs one can perform standard measurements on a joint system consisting of an original system and an ancilla in a known state. The idea of performing joint measurements of incompatible observables with the help of POVMs was discussed in the literature before -- see e.g. \cite{HRS,SRH,Busch2,Yu}.


{\bf Quantum walk algorithm for an arbitrary set of rank 1 POVM elemts.}
The following algorithm was presented in \cite{QWPOVM}: (1) Initiate the quantum walk at position $x=0$ with the coin state corresponding to the qubit state one wants to measure. (2) Set $i := 1$. (3) While $i < n$ do the following: (a) apply coin operation $C_i^{(1)}$ at position $x = 0$ and identity elsewhere and
then apply translation operator $T$; (b) apply coin operation $C_i^{(2)}$ at position $x = 1$, $NOT$ at position $x = -1$ and identity elsewhere and then apply translation operator $T$; (c) set $i := i + 1$. The generated POVM elements depend only on the choice of $C_i^{(1)}$ and $C_i^{(2)}$. This algorithm can be easily modified to include generation of rank 2 POVM elements \cite{QWPOVM}. For an exact form of coin operators used in the experiment see the supplementary material.


{\bf Joint measurement of two observables.}
The four rank 1 POVM elements for a joint measurement of noisy $\sigma_k$ and noisy $\sigma_l$ are of the form
\begin{equation}\label{}
\begin{split}
 &M_{a,b}^{k,l}=\frac{1+(a\cdot\sigma_k+b\cdot\sigma_l)/\sqrt{2}}{4}\\
 &(k, l=x, y, z; a, b=\pm1).
\end{split}
\end{equation}
We implement them using quantum walks -- for details see supplementary material. The measurement of a photon at position $x=5$ corresponds to $M_{+,+}^{k,l}$, $x=3$ to $M_{-,-}^{k,l}$, $x=1$ to $M_{+,-}^{k,l}$, and $x=-1$ to $M_{-,+}^{k,l}$.

In figure \ref{fig2}(c), BD1 and BD2, BD3 and BD4 must be aligned. By constructing  two conventional  interferometers, we observed that the interference visibilities of the interferometers are both above $0.99$. By adjusting the wave plates in the optical network, we make all the input photon pass through the output ports -1, 1, 3, and 5 respectively. Then the corresponding coupling efficiencies of single mode fibers ($\eta_1$) and the detection efficiencies of APDs ($\eta_2$) are calibrated to make sure the differences of $\eta_1\times\eta_2$  for each output port are below $5\%$.


{\bf Derivation of the relation for three unbiassed observables.}
For the three unsharp Pauli observables $X'$, $Y'$ and $Z'$, with the corresponding efficiencies $\eta_x$, $\eta_y$ and $\eta_z$, we get
\begin{eqnarray}
& & \Delta(X_{\rho},X'_{\rho})^2 + \Delta(Y_{\rho},Y'_{\rho})^2 + \Delta(Z_{\rho},Z'_{\rho})^2    \\
&=& 2\left(|\mathbf{r}\cdot \mathbf{x}(1-\eta_x)|+|\mathbf{r}\cdot \mathbf{y}(1-\eta_y)|+|\mathbf{r}\cdot \mathbf{z}(1-\eta_z)|\right). \nonumber
\end{eqnarray}
Note that $\mathbf{r}=r(\sin\theta \cos\varphi,\sin\theta \sin\varphi,\cos\theta)$ and without loosing generality we can focus on the case $\theta \in [0,\frac{\pi}{2}]$ and $\varphi \in [0,\frac{\pi}{2}]$. The above expression becomes
\begin{equation}
2r\left( \sin\theta \cos\varphi (1-\eta_x) + \sin\theta \sin\varphi (1-\eta_y) + \cos\theta (1-\eta_z) \right)
\end{equation}
, which is bounded from below by $2r(1-\max\{\eta_x,\eta_y,\eta_z\})$. Therefore, one obtains the following relation
\begin{eqnarray}
& & \Delta(X_{\rho},X'_{\rho})^2 + \Delta(Y_{\rho},Y'_{\rho})^2 + \Delta(Z_{\rho},Z'_{\rho})^2    \label{} \\
&\geq& 2r(1-\max\{\eta_x,\eta_y,\eta_z\}) \nonumber \\
&=& r \min\{ \Delta(X,X')^2,\Delta(Y,Y')^2,\Delta(Z,Z')^2\}. \nonumber
\end{eqnarray}


{\bf Joint measurement of three observables.}
The eight rank 1 POVM elements for joint measurement of noisy $\sigma_x$, $\sigma_y$ and $\sigma_z$ are given as follows:
\begin{equation}\label{}
\begin{split}
 &M_{a,b,c}^{x,y,z}=\frac{1+(a\cdot\sigma_x+b\cdot\sigma_y+c\cdot\sigma_z)/\sqrt{3}}{8}\\
 &(a, b, c =\pm1).
\end{split}
\end{equation}
Details of their quantum walk implementation are given in supplementary material. The measurement of photon at position $x=13$ corresponds to $M_{+,+,+}^{x,y,z}$, $x=11$ to $M_{-,-,-}^{x,y,z}$, $x=9$ to $M_{+,+,-}^{x,y,z}$, $x=7$ to $M_{-,-,+}^{x,y,z}$, $x=5$ to $M_{+,-,+}^{x,y,z}$, $x=3$ to $M_{-,+,-}^{x,y,z}$, $x=1$ to $M_{+,-,-}^{x,y,z}$, and $x=-1$ to $M_{-,+,+}^{x,y,z}$.

To align the six pairs BDs (BD1 and BD2, BD3 and BD4, BD5 and BD6, BD7and BD8, BD9 and BD10, BD11 and BD12), we construct the conventional interferences which interference visibilities are all above $0.99$. These ensure a good alignment. As the quantum walk step number growing, the unmounted fixed-angle wave plates are inserted into specific positions to implement corresponding coin operators. To calibrate interferences we replaced the configuration Q'HQ by a single HWP to decrease the effect of wave plates' imperfection, e.g. phase retarder and angle. Finally, all the coupling efficiencies of single mode fibers and the detection efficiencies of APDs are calibrated.

\section{Acknowledgements}
The authors thank F. L. Xiong, X. M. Hu and Z. B. Hou for useful discussions. the National Natural Science Foundation of China (grant nos. 61108009, 61222504 and 11574291). P.K. is supported by the National Research Foundation and Ministry of Education in Singapore.


\begin{thebibliography}{99}

\bibitem{Ozawa1}
Ozawa, M. Universally valid reformulation of the Heisenberg uncertainty principle on noise and disturbance in measurement. {\it Phys. Rev. A} {\bf 67}, 042105 (2003).

\bibitem{Ozawa2}
Ozawa, M. Uncertainty relations for noise and disturbance in generalized quantum measurements. {\it Ann. Phys.} {\bf 311}, 350-416 (2004).

\bibitem{Branciard}
Branciard, C. Error-tradeoff and error-disturbance relations for incompatible quantum measurements. {\it Proc. Natl. Acad. Sci. U.S.A.} {\bf 110}, 6742-6747 (2013).

\bibitem{BLWpq}
Busch, P., Lahti, P. \& Werner, R. F. Proof of Heisenberg's error-disturbance relation. {\it Phys. Rev. Lett.} {\bf 111}, 160405 (2013).

\bibitem{BLWQ}
Busch, P., Lahti, P. \& Werner, R. F. Heisenberg uncertainty for qubit measurements. {\it Phys. Rev. A} {\bf 89}, 012129 (2014).

\bibitem{Erhart}
Erhart, J. {\it et al}.  Experimental demonstration of a universally valid error-disturbance uncertainty relation in spin measurements. {\it Nature Phys.} {\bf 8}, 185-189 (2012).

\bibitem{Rozema}
Rozema, L. A. {\it et al}. Violation of Heisenberg's measurement-disturbance relationship by weak measurements. {\it Phys. Rev. Lett.}  {\bf 109}, 100404 (2012).

\bibitem{BKOE}
Baek, S. Y., Kaneda, F., Ozawa, M. \& Edamatsu, K. Experimental violation and reformulation of the Heisenberg's error-disturbance uncertainty relation. {\it Sci. Rep.} {\bf 3}, 2221 (2013).

\bibitem{Weston}
Weston, M. M., Hall, M. J. W., Palsson, M. S., Wiseman, H. M., \& Pryde, G. J. Experimental test of universal complementarity relations. {\it Phys. Rev. Lett.} {\bf 110}, 220402 (2013).

\bibitem{RBBFBW}
Ringbauer, M. {\it et al.} Experimental joint quantum measurements with minimum uncertainty. {\it Phys. Rev. Lett.} {\bf 112}, 020401 (2014).

\bibitem{KBOE}
Kaneda, F., Baek, S. Y., Ozawa, M. \& Edamatsu, K. Experimental test of error-disturbance uncertainty relations by weak measurement. {\it Phys. Rev. Lett.} {\bf 112}, 020402 (2014).

\bibitem{Entropic}
Sulyok, G. {\it et al.} Experimental test of entropic noise-disturbance uncertainty relations for spin-1/2 measurements. {\it Phys. Rev. Lett.} {\bf 115}, 030401 (2015).

\bibitem{LW}
Lund, A. P. \& Wiseman, H. M. Measuring measurement-disturbance relationships with weak values. {\it New J. Phys.} {\bf 12}, 093011 (2010).

\bibitem{BLW}
Busch, P., Lahti, P. \& Werner, R. F. Colloquium: Quantum root-mean-square error and measurement uncertainty relations. {\it Rev. Mod. Phys.} {\bf 86}, 1261 (2014).

\bibitem{BS}
Busch, P. \& Stevens, N. Direct tests of measurement uncertainty relations: What it takes. {\it Phys. Rev. Lett.} {\bf 114}, 070402 (2015).

\bibitem{Rudolph}
Korzekwa, K., Jennings, D. \& Rudolph, T. Operational constraints on state-dependent formulations of quantum error-disturbance trade-off relations. {\it Phys. Rev. A} {\bf 89}, 052108 (2014).

\bibitem{HRS}
Heinosaari, T., Reitzner, D. \& Stano, P. Notes on joint measurability of quantum observables. {\it Found. Phys.} {\bf 38}, 1133-1147 (2008).

\bibitem{SRH}
Stano, P., Reitzner, D. \& Heinosaari, T. Coexistence of qubit effects. {\it Phys. Rev. A} {\bf 78}, 012315 (2008).

\bibitem{Busch2}
Busch, P. \& Schmidt, H. J. Coexistence of qubit effects. {\it Quantum Information Processing} {\bf 9}, 143-169 (2010).

\bibitem{Yu}
Yu, S., Liu, N. L., Li, L. \& Oh, C. H. Joint measurement of two unsharp observables of a qubit. {\it Phys. Rev. A} {\bf 81}, 062116 (2010)

\bibitem{QWPOVM}
Kurzy\'nski, P. \& W\'ojcik, A. Quantum walk as a generalized measuring device. {\it Phys. Rev. Lett.} {\bf 110}, 200404 (2013).

\bibitem{Us}
Zhao, Y. Y. {\it et al}. Experimental realization of generalized qubit measurements based on quantum walks. {\it Phys. Rev. A} {\bf 91}, 042101 (2015).

\bibitem{Xue}
Bian, Z. {\it et al.} Realization of single-qubit positive-operator-valued measurement via a one-dimensional photonic quantum walk. {\it Phys. Rev. Lett.} {\bf 114}, 203602 (2015).

\bibitem{Aharonov}
Aharonov, Y., Davidovich, L. \& Zagury, N. Quantum random walks. {\it Phys. Rev. A} {\bf 48}, 1687 (1993).

\bibitem{QHQ}
Simon, B. N., Chandrashekar, C. M. \& Simon, S. Hamilton's turns as a visual tool kit for designing single-qubit unitary gates. {\it Phys. Rev. A} {\bf 85}, 022323 (2012).

\end{thebibliography}
\end{document}